\documentclass{aastex}
\usepackage{emulateapj5}
\begin{document}
\title{Polarization of GRB by Scattering off Relativistically Moving Material: Compton Sailing and High Polarization}
\author{Amir Levinson\altaffilmark{1,2}and David Eichler\altaffilmark{3}}
\altaffiltext{1}{School of Physics \& Astronomy, Tel Aviv
University, Tel Aviv 69978, Israel; Levinson@wise.tau.ac.il}
\altaffiltext{2}{School of Physics, University of Sydney, NSW 2006}
\altaffiltext{3}{Physics Department, Ben-Gurion University,
Beer-Sheva 84105, Israel; eichler@bgumail.bgu.ac.il}
\begin{abstract}
The polarization of gamma ray emission scattered off the baryon
rich material that collimates a GRB fireball and the coasting
speed of the irradiated matter are calculated numerically for
different geometries of the radiation source and the collimating
wall.  It is shown that when the scattering material is Compton
sailing, the direction of maximum polarization is quite generally
well within the beaming cone of scattered radiation. As a result,
the probability for observing bright highly polarized GRB's
increases considerably, provided the Lorentz factor of the
coasting matter is not well below 30, and the scattered radiation
is highly polarized even when the beam thickness is large compared
to $1/\Gamma$. It is suggested that correlation between
polarization and intensity could provide clues as to whether
energy flows from matter to photons or the reverse.

\end{abstract}

\section{Introduction}

 Gamma ray bursts (GRB's) are believed to contain
about $10^{51}$ ergs in the form of  several $ \times 10^{57}$
photons averaging  several hundred KeV in energy. On the other
hand, they are believed to have Lorentz factors of at least
$10^2$, and, hence, at most $\sim 10^{51}$ protons. For each
proton, there must be at least $10^5$ photons/baryon emitted.  The
question arises: what determines the number of photons emitted per
baryon and  why should the average photon energy in a GRB be
consistently close to  the electron rest mass?

A popular model posits that these photons are emitted as
synchrotron radiation in an optically thin environment, typically
at distances $10^{13}$ to $10^{15}$ cm. However, this begs the
question of why a) the synchrotron frequency in the lab frame,
which depends on the magnetic field, b) the comoving Lorentz
factor $\gamma'$ of the emitting electrons, c) the bulk Lorentz
factor $\Gamma$, and d) the viewing angle together so frequently
conspire to put the peak luminosity near 300 KeV. The fact that
this peak energy has been observed to vary from $\sim 2$ MeV all
the way down to $\sim 10$ KeV notwithstanding, we contend that the
typical value of several hundred KeV is remarkable. For example,
the strong linear correlation in the HETE II data set (Atteia et
al. 2003) between the square root of the isotropic equivalent
fluence $E_{iso}$ and the spectral peak $h\nu_{peak}$,
$h\nu_{peak}/100 KeV \sim [E_{iso}/10^{52} \rm{erg}]^{1/2}$,
suggests that the softer GRB's are intrinsically weaker and quite
possibly merely  the peripheries of GRB jets.
The bright bursts show  remarkably little variation in the
location of the spectral peak.

A  detailed analysis by Preece et al. (2002) shows that even if
the spectral peak distribution could be arranged within the
optically thin synchrotron model, the spectra at low energy, which
typically show $dn_{\gamma}/dE \propto E^{-1}$, are inconsistent
with optically thin, efficiently cooling synchrotron emission
which would give $dn_{\gamma}/dE \propto E^{-3/2}$. (It could be
argued that the bulk Lorentz factor $\Gamma$ is frequently high
enough that the synchrotron cooling is inefficient, which would
give a harder spectrum, but then it becomes hard to understand how
the emission, which depends on a high power of $\Gamma$  in this
regime, is frequently so efficient.) The recent claim of high
polarization in GRB's (Coburn and Boggs 2003) has nevertheless
strengthened the views of some that the emission mechanism is
optically thin synchrotron  (e.g. Lyutikov, Parlev and Blandford
2003). Others have argued that the claimed degree of polarization,
70 to 80 percent (but with considerable experimental uncertainty),
is too high to be accommodated by this mechanism, which limits the
polarization to at most 56 percent for the case of  a wide beam,
$\Delta\theta \gg1/\Gamma$.

At the time of this writing, the observational situation is, in
the opinion of the authors, unresolved on several fronts. The
significance of the X-ray flashes is not yet fully understood. The
degree of polarization estimated by Coburn and Boggs may be
considered uncertain by several factors: e.g. large error bars,
the chance proximity of the modulation frequency of the RHESSI
detector and the intrinsic variability of GRB021206, including  2s
variability in the hardness (Hajdas et al. 2003), and the absence
of other bursts to date that could establish how typical this
burst was. Much of the motivation for the model discussed in this
paper is thus tentative, but it should be clear that working out
the predictions theoretically is part of the process of resolving
the issues observationally.

An alternative model has been developed in a series of papers
(Levinson and Eichler, 1993, 2000, 2003; Eichler and Levinson
1999, 2000, 2003; Eichler  1994, 2003; van Putten and Levinson
2002, 2003) originally and still motivated by the above questions.
Its basic assumption is that the extreme baryon purity that must
exist in the high $\Gamma$  GRB outflows is enforced by the  event
horizon of a central black hole, so that the original baryonic
content in the outflow must vanish. The photosphere near the axis
of the outflow is therefore at about the annihilation radius of
the pairs, which until then constitute the main component of the
fireball plasma.  The fireball's ability to manufacture more
photons drops precipitously beyond the e$^{\pm}$ annihilation
radius, so the total
photon entropy of the GRB is frozen near its value at pair
annihilation, when the average energy per photon is of the order
of magnitude of $m_ec^2$. [Actually, the comoving temperature
should be about an order of magnitude  (the log of the compactness
parameter)  less than $m_ec^2$ if there is perfect thermal
equilibrium, but the dissipation introduced by the friction of the
fireball with the surrounding walls probably introduces some
non-equilibrium signature on the spectrum, and the bulk Lorentz
factor is probably at least a few even in the presence of such
friction.]

In this class of models, the average photon energy in a GRB,
though somewhat dependent upon the geometry of the walls and how
the photons bounce off them,  is naturally of order $m_ec^2$
without fine tuning of the parameters. As long as photons can
bounce off the walls at large angles and make pairs with other
outgoing photons, the dissipation continues,
 the expansion is accompanied by additional photon production, and  the bulk Lorentz
factor remains moderate. As soon as the energy density is below
that needed to maintain pairs, transparency sets in and the photon
entropy remains nearly constant thereafter.

Yet another class of models posits that GRB's are photons that are
upscattered optical/UV photons by relativistic electrons with a
Lorentz factor of order 300 (Shaviv and Dar 1995a, b; Lazzati et
al. 2000; Dado, Dar, and De Rujula 2003).  In these models, the
number of photons is conserved and the energy of the outgoing
photons, which is provided by the Compton scattering material, has
been increased by a factor of $10^5$ to $10^6$. Thus, the once
scattered photons constitute the primary GRB beam. Because the
number of photons is conserved, the isotropic equivalent
luminosity of the Compton scattered photons is  of order
$\Gamma^4$ times that in the optical/UV photons. If the latter is
of order the Eddington luminosity for a $10 M_{\odot}$ star, then
the observed isotropic luminosities for bright bursts, $10^{13}$
to $10^{15}L_{edd}$ require Lorentz factors exceeding $10^3$.
Shaviv and Dar (1995) predicted in their model that inverse
Compton emission is expected to emerge polarized with a
correlation between polarization and other quantities associated
with the angle of observation.  This should include overall
fluence and peak luminosity, which are down from the peak values
by factors of 8 and 16 respectively when the polarization is at a
maximum. The high polarization is possible only when the angular
width of the beam is narrower than $1/\Gamma$, and this constrains
the solid angle of the beam to be extremely small if high net 
polarization is to be observed.

The Compton upscattering model for polarization (Dar and Shaviv
1995) should be compared and  contrasted to the ensheathment model
of Eichler and Levinson (1999). Both models predicted linear
polarization before the announced discovery (Coburn and Boggs
2003) In the ensheathment model, the energy of the photon is not
necessarily increased by the scattering and certainly not by a
dramatic factor. The photons can thus originate in a much smaller
region without violating the black body limit, and short
timescales are, in principle, possible even for the once scattered
radiation (the criticism of Lazzati et al. on this point is
therefore not justified). The   scattering material in the
ensheathment model may have a relatively moderate bulk Lorentz
factor and the solid angle of the scattered light can subtend a
much larger angle than the high $\Gamma$ scattering material
invoked by the Compton upscattering models. It is therefore
possible to combine a large (lab frame) scattering angle
and high polarization.  It is also easy in the ensheathment model
to have a geometry in which only scattered radiation is seen by
the observer. As proposed by Eichler and Levinson (1999), the
large reduction of the "on-axis" luminosity that was apparently
achieved by GRB980425/SN1998bw, perhaps 4 or 5 orders of
magnitude, comes with only modest reduction of frequency.  It can
result from a small fraction of photons being scattered, by a
non-relativistic or modestly relativistic scatterer, into such a
large angle as to be observable by us. The smooth light curve that
was observed for GRB980425, rare among GRB's and far removed from
the relation of Atteia et al. (2003), is consistent with the light
having reached us via many different paths.

 In recent years, the view that GRB's are  collimated
 (e.g. Levinson and Eichler 1993)has been
almost universally accepted, and the collimating matter is
probably a host star or a wind. This confirms the assumption that
GRB fireballs pass through some sort of surrounding wall in their
early stages. This has particular significance for models in which
the early fireball is baryon-free, because it provides a source of
baryon seeding that could account for afterglow-generating plasma
escaping intact to infinity.  In the first place, some baryons are
introduced downstream as entrained matter or pick-up neutrons that
diffuse into the outflow from surrounding walls of baryonic
matter. This matter can be a host star or simply a baryonic wind
driven from the accretion disk of the black hole. The baryonic
density in this picture has a complicated transverse structure
(Levinson and Eichler 2003), the density increases outward, and
the photosphere can be close to the pair annihilation radius near
the axis, about $10^{10}$ cm, and near the walls at the edge (see
Fig. 1).

In this paper, we  consider another source of baryons - the
collimating walls themselves  - and note that, due to the impact
of the GRB photons  on the walls, relativistic motion of the walls
can ensue. This matter can then be considered part of the GRB. It
can cause highly polarized gamma ray emission in certain viewing
directions and can cause afterglow. It might even be the case that
there is a smooth transition from the supernova ejecta to  the GRB
fireball. This could account, for example, for the peculiar
features of SN 1998bw, which has a huge expansion velocity (c/6)
and peculiar abundances.

This was considered by Eichler (2003, hereafter E03) under
simplifying geometric assumptions: that the photons emerged from a
small, effective point source.   It was argued that the photon
pressure would accelerate the  wall material to a Lorentz factor
of $1/sin\chi$, where $\chi$ is the angle between the photon flow
and the wall, and that the polarization of scattered photons off
the wall would be greater than 0.6 with considerable probability.
Here we relax the point source assumption. In section 2 we
consider the terminal Lorentz factor in the presence of a photon
field from a finite source, whose shape is characterized by two
parameters.  In section 3, we consider the effect of finite source
size on the expected polarization. Finally, we consider multiple
scattering within a localized neighborhood of the point of photon
impact on the sheath.

We demonstrate that in the present model, large polarization (of
order 40 percent) is typical even when the beam "acausal", i.e.
when it has a much larger opening angle than $1/\Gamma$. This is
not the case in most models where the intensity along the line of
sight typically receives contributions from a surrounding
neighborhood of velocity vectors within the beam, whereby the
polarizations of the various contributing beams cancel.  In the
limit of axisymmetric contribution around the line of sight, this
cancellation would be complete.  In the present model, on the
other hand, there is a lower degree of symmetry (hence the
calculations are unfortunately more complicated), because the
incident radiation arrives from a direction much different (even
in the comoving frame) from the velocity vector.

\section{Polarization by a Compton Sailing Sheath}

We consider the dynamics of the inner layers of the baryon rich
sheath surrounding the baryon poor core, and the consequent
polarization of the scattered gamma-ray emission.  We envision
that the sheath is irradiated by the gamma-ray source which is
located at small radii, as illustrated in Fig 1.  The radiative
force would accelerate the inner layers, leading to a strong shear
of the boundary layer.  Viscous forces may counteract the
radiative force, and the profile of the Lorentz factor across the
boundary layer would, in general, depend on the detail of momentum
transfer between the moving layers of the sheath. In the limit of
a highly super-Eddington primary photon luminosity,  it seems
likely that the radiation pressure overwhelms the viscous forces
on the exposed layers of scattering material. This case has been
considered in E03, and is explored further below.

\subsection{Dynamics}
The stress-energy tensor of a magnetized flow is given by
\begin{equation}
T^{\alpha\beta} = hnU^{\alpha}U^{\beta} - p\eta^{\alpha\beta}+\frac{1}{4\pi}
(F^{\alpha\sigma}F^{\beta}_{\sigma}+\frac{1}{4}\eta^{\alpha\beta}F^2),
\label{Tmn}
\end{equation}
where $F_{\mu\nu}$ is the electromagnetic tensor,
$U^\alpha=(\Gamma,\Gamma\vec{\beta})$ is the 4-velocity of the bulk
fluid, and $n$, $p$, and $h$, are the proper particle density,
pressure and specific enthalpy, respectively.  The energy,
momentum and continuity equations can be written as

\begin{eqnarray}
\partial_{\alpha}T^{\alpha\mu}=S^{\mu},\nonumber \\
\label{eq: MHD}
\partial_{\alpha}(nU^{\alpha})=0.
\end{eqnarray}
The above set of equations must be augmented by Maxwell equations
for the electromagnetic field. The source term $S^\mu$ accounts
for all external forces acting on the MHD flow, and is expressed
here as the sum of a radiative force and all other forces (e.g.,
viscosity mediated by some instabilities, electromagnetic
stresses, or diffusing neutrons; see Levinson and Eichler 2003):
$S^\mu=S_c^\mu+S_2^\mu$. In terms of the distribution functions of
gamma rays, $f_\gamma$, and electrons (we don't distinguish here
between electrons and positrons), $f_e$, the source term
associated with the radiative force is given, in the limit of
Thomson scattering, by (see e.g., Phinney 1982),

\begin{equation}
S_c^\mu=-c\sigma_T\int{\frac{d^3p_e}{u^0}\int{\frac{d^3p_\gamma}{p_\gamma^0}f_\gamma
f_e u_{\alpha}p_\gamma^\alpha[p_\gamma^\mu+(u_{\nu}p_\gamma^\nu)u^\mu]}},
\label{Sc}
\end{equation}
where $p_e^\mu$ and $p_\gamma^\mu$ are the 4 momenta of electrons
and gamma rays, respectively, as measured in the Lab frame, and
$u^\mu=p_e^\mu/m_e$ is the corresponding electron 4 velocity. The
gamma-ray distribution function satisfies a Boltzman equation:
\begin{equation}
p_\gamma^\mu\partial_{\mu}f_\gamma(p_\gamma,x)=C_\gamma(f_\gamma,p_\gamma,x),
\label{Boltz}
\end{equation}
where the collision operator must satisfy, $\int{p_\gamma^\mu
C_\gamma d^3p_\gamma=-S_c^\mu}$, by virtue of energy and momentum
conservation.  Once $S_2^\mu$ is specified, the above set of
equations, subject to appropriate boundary conditions, can, in
principle, be solved to yield the structure and dynamics of the
boundary layer. A complete treatment will be presented elsewhere.

In what follows, we consider, for illustration, the case of a
sufficiently cold fluid dominated by radiative forces. By
sufficiently cold we mean that to a good approximation the
electron distribution function is given by $f_e=\Gamma
n_e\delta^3(p_e^i-m_eU^i)$, where $n_e$ is the electron proper
density. We further suppose that the system is stationary and
axially symmetric. Projecting eq. (\ref{eq: MHD}) on the direction
perpendicular to the wall, defined by the unit vector $\hat{n}_s$,
and using eq. (\ref{Tmn}), yields the pressure gradient required
to support the corresponding component of the radiative force:
\begin{equation}
\hat{n}_s\cdot\vec\nabla(p+B^2/8\pi)=-\frac{\sigma_T
n_e\Gamma}{c}\int{d\nu} \int{d\Omega
I_{\nu}(1-\beta\mu)\sin\theta}.
\label{press}
\end{equation}
Here, $I_\nu=p_\gamma^3hcf_\gamma$ is the gamma ray intensity, and
$\mu=\cos\theta$ denotes the cosine of the angle between the
direction of impinging photons and the fluid velocity. The zeroth
component of eqs. (\ref{eq: MHD}) yields

\begin{eqnarray}
\vec{\beta}\cdot\vec\nabla(h\Gamma+B^2/4\pi n)=-\frac{n_e\sigma_T}{n c}\times\cr \int{d\nu}
\int{d\Omega} I_{\nu}(\Omega)(1-\beta\mu)&[\Gamma^2(1-\beta\mu)
-1],
\label{force}
\end{eqnarray}
where the continuity equation has been used.

The intensity at a given point on the wall generally receives
contribution from both  direct illumination of the radiation
source and from reflection from other parts of the sheath that are
exposed to the radiation source. To calculate the intensity we
must therefore solve a transfer equation that accounts for
multiple scattering between the different parts of the wall, as
well as for diffusion of photons outward into the denser parts of
the sheath. Since the reflected emission is beamed due to the
relativistic motion of the sheath, it is clear that eq.
(\ref{force}) and the transfer equation obeyed by the intensity
are coupled. The solution of the radiative transfer equation is
beyond the scope of this paper.  To simplify the analysis, we
consider, in what follows, direct irradiation of the wall by an
effective radiation source of a finite extent. From eq.
(\ref{force}) it is evident that for sufficiently large incidence
angles of the impinging photons, the fluid will quickly
accelerates to its coasting speed at which the component of the
radiative force in the direction of motion nearly vanishes (E03).

In order to calculate the Lorentz factor of the coasting layer, we
suppose that the incident radiation is emitted from a surface
defined by $S=S(x,y,z)$.  The coordinate system adopted is shown
in Fig. 1.  The $z$ axis is chosen to coincide with the symmetry
axis. We denote by $\vec{r}_\beta$ the position vector of a fluid
element, and by $\vec{r}_\gamma$ the position vector of a point on
the gamma-ray emitting surface.  The direction of an incident
photon emitted from a point defined by $\vec{r}_\gamma $ is then
given by (see Fig. 1)
\begin{equation}
\hat{k}_i= \frac{\vec{r}_\beta-\vec{r}_\gamma}{|\vec{r}_\gamma-\vec{r}_\beta|}.
\label{kin}
\end{equation}

The geometry of the radiation source adopted in the calculations
presented below is shown in Fig. 2.  The gamma-ray emitting
surface has a geometry of a ring with inner and outer radii
denoted by $a_1$ and $a_2$, respectively.  The ring lies in the
$(x,y)$ plane at $z=0$.  The radius of the wall at $z=0$ is
denoted by $d$, and is normalized such that $d=1$.  The baryonic
wall is taken to be conical for convenience.  We note, however,
that the difference between a conical wall and a curved one (as in
the case of collimation) is only in the definition of $d$.  In the
latter case $d$ represents the radius of a fictitious cone that
extends from the location of the scattering material along the
velocity vectors.  The intensity in the source is taken to be
homogeneous, that is, independent of $x$ and $y$ inside the ring
(and zero outside, of course). The spectrum of  incident radiation
is taken to be a power law with index $\alpha=0.5$ ($I_{in}\propto
\nu^{-\alpha}$). The ring emission is allowed to be beamed.  The
beaming is parametrized by an angle $\theta_{\gamma}$, measured
with respect to the symmetry ($z$) axis. For a given beaming angle
$\theta_\gamma$, only rays that satisfy the condition
$\hat{k}_{i}\cdot\hat{z}=\cos(\theta_\gamma)$ for the $\hat{k}_i$
given by eq. (\ref{kin}) can scatter off a fluid element located
at $\vec{r}_\beta$.

In the coasting regime the R.H.S of eq.  (\ref{force}) nearly vanishes, and
the solution of eq. (\ref{force}) reduces to
\begin{equation}
\beta=\kappa-(\kappa^2-1)^{1/2},
\label{beta}
\end{equation}
with
\begin{equation}
\kappa=\frac{\int d\nu\int{d\Omega_s I_\nu (1+\mu^2)}}{2\int d\nu\int{d\Omega_s \mu I_\nu}},
\label{kappa}
\end{equation}
where the inner integration is carried over the gamma-ray emitting surface, $S$, with
$d\Omega_s=(d\hat{S}\cdot\vec{r}_\beta/r_\beta^3)$.

In the limit $a_1=a_2=0$ that corresponds to a point source
located on the symmetry axis, the intensity is given by
$I\propto\delta(\mu-\mu_0)$, with $\mu_0=z/(z^2+d^2)^{1/2}$, and
eq. (\ref{beta}) reduces to $\beta=\mu_0$, recovering the result
obtained by E03.

\subsection{Polarization}

Consider now the polarization of the radiation scattered by the moving wall. Using the coordinate
system depicted in Fig.1, we express the velocity vector of a fluid element, measured with
respect to the Lab frame, as
$\vec{\beta}=\beta(\sin\theta_\beta\cos\phi_\beta\hat{x}
+\sin\theta_\beta\sin\phi_\beta\hat{y}+\cos\theta_\beta\hat{z})$, and the line of sight direction
as $\hat{n}=\sin\theta_n\cos\phi_n\hat{x}+\sin\theta_n\sin\phi_n\hat{y}+\cos\theta_n\hat{z}$.
The vector $\hat{k}_i$ given by eq. (\ref{kin}), and the vector $\hat{n}$ are transformed into
the comoving frame of the fluid element (henceforth denoted by prime), using the aberration
formula:
$\cos\theta^\prime_{i\beta}\equiv\hat{k}^\prime_i\cdot\hat{\beta}
=(\hat{k}_i\cdot\hat{\beta}-\beta)/[1-\beta(\hat{k}_i\cdot\hat{\beta})]$,
and likewise for $\cos\theta_{n\beta}^\prime=\hat{n}^\prime\cdot\hat{\beta}$.

Let $I_{in}(\vec{r}_\gamma,\hat{k}_i,\nu)$, with $\hat{k}_{i}$ given by eq. (\ref{kin}),
denote the intensity of the incident radiation.  In the rest frame of the fluid element
it is given by  $I^\prime_{in}(\vec{r}^\prime_\gamma,\hat{k}^\prime_i,\nu^\prime)=
\delta_{i}^3I_{in}(\vec{r}_\gamma,\hat{k}_i,\nu)$, where
$\delta_i=\nu^\prime/\nu=[\Gamma(1+\beta\cos\theta^\prime_{i\beta})]^{-1}$
is the corresponding Doppler factor.
The polarization vector in the comoving frame lies along the direction
$\hat{t}=\hat{n}^\prime\times\hat{k_i}^\prime/|\hat{n}^\prime\times\hat{k_i}^\prime|$.
In terms of the unit vector
$\hat{e}=\hat{n}^\prime\times\hat{\beta}/|\hat{n}^\prime\times\hat{\beta}|$,
the Stokes angle is
$\cos\chi=\hat{t}\cdot\hat{e}$.
The comoving emissivity is given by
\begin{equation}
j_{\beta}^\prime=\frac{3}{16\pi}\sigma_T n_e^\prime\int{ I_{in}^\prime(1+\cos\psi^{\prime
2})d\Omega_s},
\label{jbeta}
\end{equation}
and the corresponding Stokes parameters by
\begin{equation}
\left(
\begin{array}{c}
q_\beta^\prime\\
u_\beta^\prime\\
\end{array}
\right)
=\frac{3}{16\pi}\sigma_T n_e^\prime\int{d\Omega_s I_{in}^\prime(1-\cos\psi^{\prime 2})}
\left(
\begin{array}{c}
\cos2\chi\\
\sin2\chi\\
\end{array}
\right),
\label{Stokbeta}
\end{equation}
where $\cos\psi^\prime=\hat{n}^\prime\cdot\hat{k_i}^\prime$.
If we now measure the Stokes angle with respect to the fixed vector $\hat{b}=
\hat{n}\times\hat{z}/|\hat{n}\times\hat{z}|$, then the Stokes parameters are
rotated by an angle $-2\eta$, where $\cos\eta=\hat{b}\cdot\hat{e}$, viz.,
\begin{equation}
\left(
\begin{array}{c}
q_\beta^\prime\\
u_\beta^\prime\\
\end{array}
\right)
\rightarrow
\left(
\begin{array}{c}
q_{\beta}^\prime \cos2\eta-u_{\beta}^\prime \sin2\eta \\
q_{\beta}^\prime \sin2\eta+u_{\beta}^\prime \cos2\eta\\
\end{array}
\right).
\label{rot}
\end{equation}
Transforming to the Lab frame and averaging over directions of emitting fluid elements,
we finally obtain
\begin{equation}
I=\frac{3}{16\pi}\int{\frac{\tau}{\Gamma}\int{ j_{\beta}^\prime \delta_n^{k}d\phi_\beta}d\theta_
\beta},
\label{I}
\end{equation}
and
\begin{equation}
\left(
\begin{array}{c}
Q\\
U\\
\end{array}
\right)
=\frac{3}{16\pi}\int\frac{\tau}{\Gamma}\int{
\left(
\begin{array}{c}
q_{\beta}^\prime \cos2\eta-u_{\beta}^\prime \sin2\eta \\
q_{\beta}^\prime \sin2\eta+u_{\beta}^\prime \cos2\eta\\
\end{array}
\right)
\delta_n^{k}d\phi_\beta d\theta_\beta},
\label{Stock}
\end{equation}
where $\tau=\sigma_T n_e l$ is the optical depth along the sight
line, and $\delta_n=[\Gamma(1-\beta\cos\theta_{n\beta})]^{-1}$ is
the corresponding Doppler factor. The factor $1/\Gamma$ is
introduced because of the transformation of the electron density
from the rest frame of the fluid element to the Lab frame.  Here
we allow for a spread in the opening angle of the conical wall.
The dependence of $n_e$ and $\Gamma$ on the angle $\theta_\beta$
should be specified in general.  Below we examine two cases: a
thin cone, defined by $n_e(\theta_\beta)\propto
\delta(\theta_\beta-\theta_0)$, and a cone with a constant density
in the region
$\theta_0-\Delta\theta\le\theta_\beta\le\theta_0+\Delta\theta$ and
zero outside this range. The index $k$ depends, in general, on the
spectrum of incident radiation, as well as on the kinematics of
the emission region, and is expected to lie in the range from 2.5
to 3.5.  The dependence of the polarization on $k$ is explored
below. Finally, the polarization degree is given by
$P=(Q^2+U^2)^{1/2}/I$.  We note that for an axially symmetric
distribution of the scattering material, the Stokes parameter $U$
should vanish (Begelman \& Sikora, 1987).  As a check, we numerically computed $U$ in each
run and verified that, within the accuracy of our computation, it
is zero.

We also calculate below the probability for observing a source with a polarization $P>\pi$,
given explicitly by,
\begin{equation}
x(P>\pi)=\frac{1}{A}\int_{P(\mu)>\pi}{V_{max}d\Omega}=\frac{2\pi}{A} \int_{P(\mu)>\pi}{I^{3/2}d\mu},
\label{x(P)}
\end{equation}
where the integration is carried over all solid angles for which the polarization fraction
$P$ exceeds the value $\pi$, and $A=\int_{P(\mu)\ge 0}{V_{max}d\Omega}$ is a normalization
constant.

\section{Results}

Eqs. (\ref{jbeta}) through (\ref{Stock}) have been integrated
numerically to yield the intensity and polarization of the
radiation scattered off a section of the wall that moves with a
given velocity.  In cases corresponding to  Compton sailing
material, the velocity of the scattering matter was first
calculated using eqs. (\ref{kappa}) and (\ref{beta}). As a check
on our code, we computed the scattered intensity for a point
radiation source located on the symmetry axis ( $a_1=a_2=0$), and
compared the results with the analytic approximation derived by
E03.  An example is shown in Fig. 3 for $z=100$, $k=3$, and
$\theta_0=0.1$.  The corresponding Lorentz factor of the Compton
sailing fluid is $\Gamma=101$.  The numerical result is given by
the dashed line and the analytic one by the solid line.  We also
plotted the polarization fraction $P$ (dotted line).  As seen, the
agreement is good.  The slight asymmetry of the dashed curve
around the direction of fluid motion, $\Delta=0$, is due to
second order effects associated with the curvature of the wall
which are not accounted for by the analysis of E03. Note also that
the viewing angle along which the intensity peaks is slightly
shifted from the direction $\Delta=0$, and in particular does not
coincide with the direction of maximum polarization. This is due
to the angular dependence of the Thomson cross section. This is
the reason for the relatively rapid decline of the $x(P>\pi)$
curve. As a second test case, we calculated the polarization for a
radiating ring of radii $a_1\rightarrow a_2=1$, with a highly
beamed emission ($\theta_\gamma<<1$). This case corresponds to the
head on approximation considered in Eichler and Levinson (2003).
The Lorentz factor of the moving wall in this case is arbitrarily
chosen (not the Compton sailing value, of course).  The resultant
polarization curves are displayed in Fig. 4. We find perfect
agreement as $\theta_\gamma\rightarrow 0$.

The dependence of the polarization produced by a Compton sailing
sheath on the various parameters is examined in Figs. 5 - 7.  The
upper panel in each figure exhibits the polarization fraction $P$
versus the line of sight angle $\theta_n$, and the lower panel the
corresponding Euclidean probabilities $x(P>\pi)$, given by eq.
(\ref{x(P)}), against $\pi$.

 Note that in the limit that sources
are so bright that they can be seen anywhere in the universe at
most viewing angles, then the probability of a given polarization
is indicated  by $\Delta \theta$, i.e. the x axis in the upper
panels of figures 5-7, except for effects from cylindrical
geometry, which are small when $\theta_o\Gamma \gg 1$ and
$\theta_o/ \Delta \theta \gg 1$. In the opposite limit, probably
not the most relevant for long, hard bursts, we assume Euclidean
geometry with $V/V_{max}=.5$. The probabilities for observing
polarization $\pi$ almost certainly lies between these two limits.

Fig. 5 shows the dependence of the polarization curves on the
opening angle of the flow, $\theta_0=\hat{\beta}\cdot\hat{z}$. In
this example $a_2=d=1$, that is, the radiating ring extends from
the wall, and $a_1=0.9$, as indicated.  The ring is assumed to
radiate isotropically in this example (more precisely,
$\theta_\gamma >2\theta_0$ is assumed). As seen, the polarization
approaches zero on the axis ($\theta_n=0$), as required by the
axial symmetry.  The Lorentz factor of the Compton sailing fluid
obtained in this run is $\Gamma\simeq 56$ for $\theta_0=0$ and ,
$\Gamma\simeq 57$ for $\theta_0=0.1$. The reason that $\Gamma$
increases slightly with increasing $\theta_n$, is that the angle
between the directions of incident photons and $\hat{\beta}$
slightly decreases on the average, but since the ring radiates
isotropically and the distance $z$ from the ring to the scattering
material is rather large, the effect is small.  The polarization
peaks at $\theta_n=\theta_0$ (but is not 100\% owing to the
contribution from elements moving at different $\phi_\beta$), as
long as $\theta_0$ is not too small.  Quite generally we find that
the maximum polarization is suppressed for sufficiently small
$\theta_0$, as seen in Fig. 5, by virtue of the axial symmetry.
The kinks in the polarizations curves correspond to a $90^\circ$
change in the polarization vector.

The dependence of the polarization on the index $k$ is displayed
in Fig. 6.  The Lorentz factor obtained for all cases shown in
this example is $57$.  As seen, the $P$ curves depend weakly on
$k$, and mainly in the side lobes, owing to the large beaming
factor of the scattered emission.  The primary reason is that the
main bump of the polarization curves in all cases lies in the
range $\Gamma\Delta<<1$.  For much smaller Lorentz factors we find
somewhat stronger dependence on $k$.  The intensity is more
sensitive to $k$, and this is reflected in the $V_{max}$ curves,
as seen.

Fig. 7 exhibits the dependence of the polarization on the geometry
of the incident radiation source.  Three cases are examined: a
point source ($a_1=a_2=0$), a disk extending from the wall
($a_1=0$, $a_2=1$), and an infinitely thin ring ($a_1=a_2=1$). As
seen, a point source gives rise to a nearly maximum polarization
in the direction $\theta_n=\theta_0$, as expected in the limit of
large beaming.  This is due to the fact that all incident rays are
perpendicular to the direction of motion in the comoving frame.
The peak of the polarization curve is smaller in the case of an
extended source by virtue of the finite range of incidence angles,
but is still rather high for sufficiently large Lorentz factors
($\Gamma=64$ in the case of the disk and $56$ for the thin ring).

In Fig. 8 we explore how the polarization depends on the Lorentz
factor of the Compton sailing matter. All three cases shown were
calculated using a point radiation source ($a_1=a_2=0$).  The
source is located at a different distance from the scattering
material in each case.  The Lorentz factors of the sailing
material, roughly equal $\Gamma=(z^2+d^2)^{1/2}/d$ are indicated.
As expected, the maximum polarization is reduced for small Lorentz
factors, owing to the larger contribution from fluid elements
located at larger angular separation with respect to the sight
line.  We found, however, that the cumulative probability curves
converge to the solid line for $\Gamma>25$, as can be inferred
from the figure.

Fig. 9 examines the effect of Compton sailing.  Here we compare
the cumulative $V_{max}$ curve obtained in the Compton sailing
case (indicated as $\Gamma_c$ in the figure) with cases obtained
when the Lorentz factor deviates from this value.  We also plotted
for comparison the result obtained in the head on approximation in
the limit $\Gamma\theta_0\rightarrow 0$ (Eichler and Levinson
2003). It is seen that Compton sailing greatly increases the
likelihood of strong polarization.

The results described above correspond to an infinitely thin cone
with an opening angle $\theta_0$. The effect of a spread in the
opening angle of the cone of optically thin scattering material is
examined in Fig. 10.  In this example the velocity vectors are
assumed to lie in the range between $\theta_0-\Delta\theta$ and
$\theta_0+\Delta\theta$.  All fluid elements are assumed to move
with their Compton sailing Lorentz factor.  As seen, the
polarization is lowered when  $\Delta\theta$ is  much larger than
$1/\Gamma$, but, in  the absence of any cylindrical symmetry
around the line of sight, does not plummet to zero in this limit.

Finally, the effect of photon diffusion in cases of a large
optical depth is depicted in Fig. 11. Solving the radiative
transfer in a medium with a modest optical depth is a formidable
task, and is beyond the scope of this paper.   Here we consider
diffuse reflection from a plane parallel, semi-infinite slab.  We
suppose that in the comoving frame the radiation incident normal
to the wall, and use the results derived in Chandrashekar (1960)
for the polarized intensity reflected off the wall. We then
transform the intensities back to the Lab frame in order to
compute the probability $x(P>\pi)$. The upper panel exhibits the
polarization curve and the (normalized) Euclidean $V_{max}$ curve,
and the lower panel the cumulative probability as usual.  The
direction of maximum polarization is along the wall of course,
with peak polarization of 0.47.  The total intensity of the
reflected radiation increases with increasing angle from the wall,
and is maximum at an angle of $1/\Gamma$, as measured in the Lab
frame (about a factor of 2 larger than along the wall).  This is
reflected in the shape of the $V_{max}$ curve.

\section{Conclusions}

We have shown, under less simplified assumptions than made in
Eichler (2003), that when matter is Compton dragged into a
"sailing" equilibrium by photons, the polarization can be quite
high for  singly scattered photons.  The direction of maximum
polarization coincides with the maximum intensity of the polarized
component. Finite source size is found to  not qualitatively alter
this conclusion. Finite source angular width $\delta \theta$,
which can nearly eliminate net polarization in most models, is
found in the present model to reduce it only somewhat, and to
leave about 40 percent polarization for most relevant viewing
angles.

In the  picture presented here,  singly scattered photons may in
principle compete with unscattered ones, because the photon energy
as seen in the lab frame  is not dramatically altered. [This is in
contrast to the scenario of Shaviv and Dar, 1995, Lazzati et al.
2000, where the photon energy is dramatically raised by the
scattering, so that virtually all the gamma rays are attributed to
such scattering.] However, the scattered and unscattered photons
do not necessarily arrive at the same time, are not concentrated
in the same directions,  and do not necessarily have exactly the
same spectrum. [In  some cases, the line of sight does not
necessarily admit unscattered photons, e.g. as was argued
(Nakamura 1998, Eichler and Levinson 1999) for the case of
GRB980425, however, this GRB may be exceptional.] It is thus important
that future GRB polarization experiments be sensitive to the
polarization as a function of time and energy range.

 If the photons are scattered (undergoing "diffuse reflection" in the
terminology of Chandrasekhar, 1960) off an optically thick sheath,
then the maximum polarization corresponds to 47 percent along the
direction of maximum intensity of the polarized component.
Typically, polarizations of 30 to 47 percent are expected in this
situation among the scattered photons. Consistently observed
values in this range of polarizations, with a sharp cutoff at 47
percent, might thus be a signature of diffuse reflection.

The situation in which the optical depth of the scatterers is
slightly greater than unity remains to be studied in detail. We
believe that the results will be not too much different from the
case of single scattering, because the path of maximal
polarization corresponds to minimum optical depth in a
relativistically expanding flow. In any case, they should lie
between the two extremes considered in this paper. It might be
claimed that an optical depth of order unity is "fine tuned".
However, in the present picture, the sheath tapers to infinity, so
most of its area is in fact comprised of material with an optical
depth of order unity. The path of least optical depth is along the
direction of motion, because then the relative velocity between
the photon and the material, as seen by the lab frame is the
smallest (Levinson and Eichler, 2003).

If future polarization studies eventually show some indication of
scattering of GRB photons, correlation between maximum
polarization and maximum intensity might provide clues as to
whether energy is indeed flowing from the scattering material to
the photons or whether most of the scattering occurs after Compton
equilibrium has been reached, which would suggest that matter is
being dragged by the photons rather than upscattering them.

This research was supported by the Israel-U.S. Binational Science
Foudnation, the Arnow Chair of Theoretical Astrophysics and Ben
gurion University, and by  an ISF grant for a Israeli Center for
High Energy Astrophysics.

\clearpage


\newpage
\begin{figure}[hb]
\centerline{\epsfxsize=140mm\epsfbox{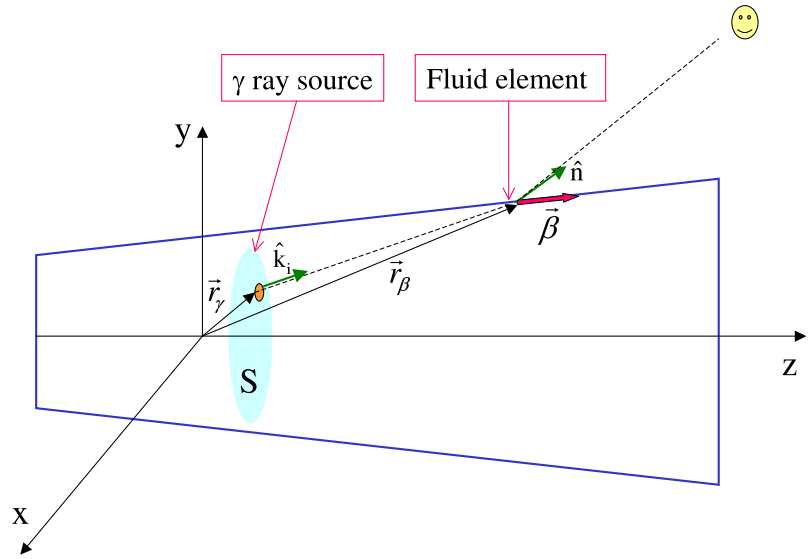}}
\caption{Schematic illustration of scattering by a relativistic sheath.
Gamma rays are emitted from a surface marked as S in the figure,
and scatter off matter that ensheath the ultrarelativistic,
optically thin core.  The symmetry axis of the system is taken to be
in the $z$-direction. The position of a scattering center on the wall,
moving with a velocity $\vec{\beta}$, is defined by the vector $\vec{r}_\beta$.
The position vector of a point on the gamma-ray emitting
surface is labeled by $\vec{r}_\gamma$.  The sight line direction is denoted by $\hat{n}$.}
\label{f1}
\end{figure}

\newpage
\begin{figure}[hb]
\centerline{\epsfxsize=140mm\epsfbox{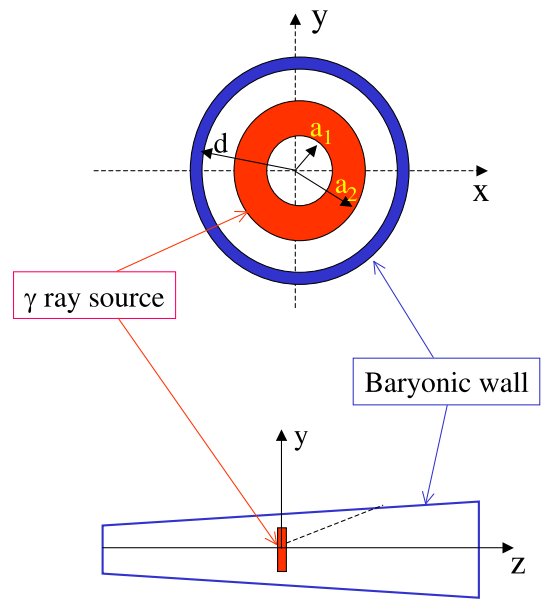}}
\caption{A sketch of the radiation source geometry adopted for the calculations is
shown.   The emitting surface has a geometry of a ring with inner and outer radii
denoted by $a_1$ and $a_2$, respectively.  The ring lies in the $(x,y)$ plane
at $z=0$ and is centered around the symmetry axis of the system.  The radius of the
wall containing the scattering material
at $z=0$ is denoted by $d$.}
\label{f2}
\end{figure}

\newpage
\begin{figure}[hb]
\centerline{\epsfxsize=140mm\epsfbox{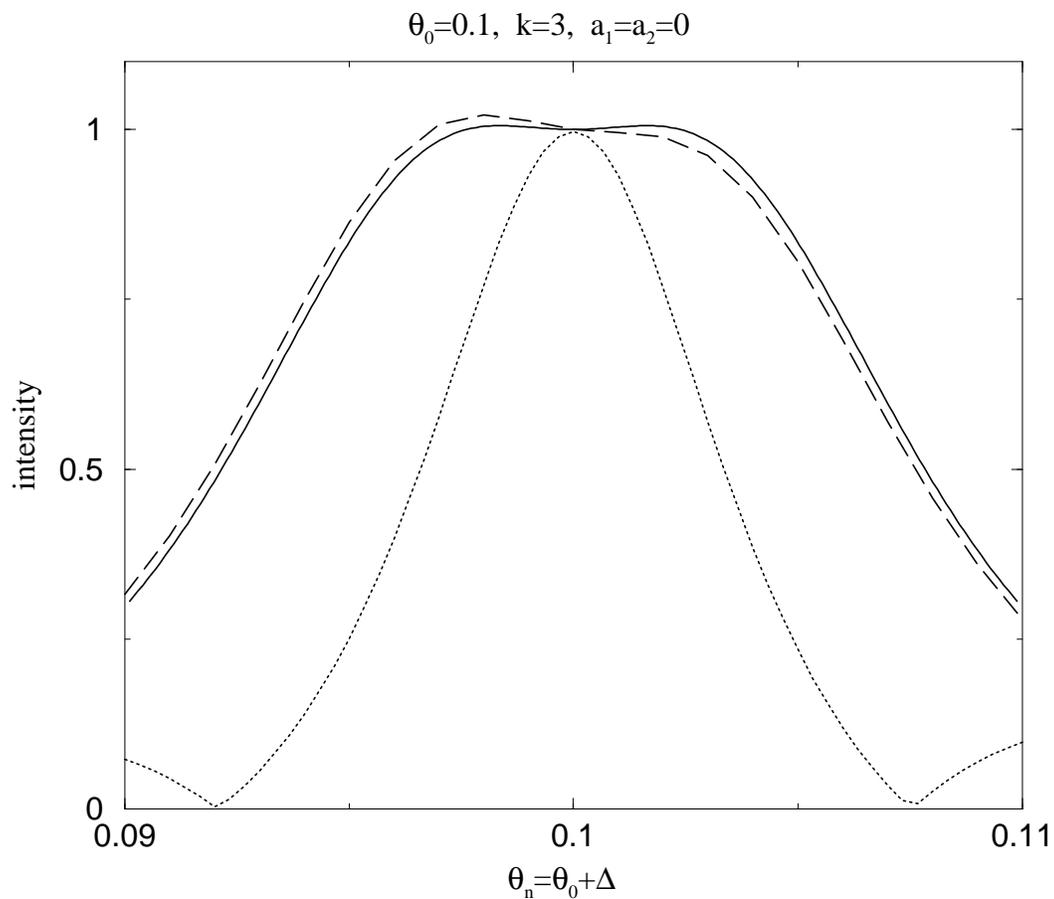}}
\caption{
Scattered intensity (dashed lines) and polarization
degree (dotted line), calculated using a point radiation source located on the
symmetry axis, are plotted as a function of viewing angle $\theta_n$.  A plot of
the intensity calculated using the analytic approximation derived in Eichler (2003)
is shown for a comparison (solid line).  The parameters $\theta_0$ and $k$ are defined
in the text.}
\label{f3}
\end{figure}
 
\newpage
\begin{figure}[hb]
\centerline{\epsfxsize=140mm\epsfbox{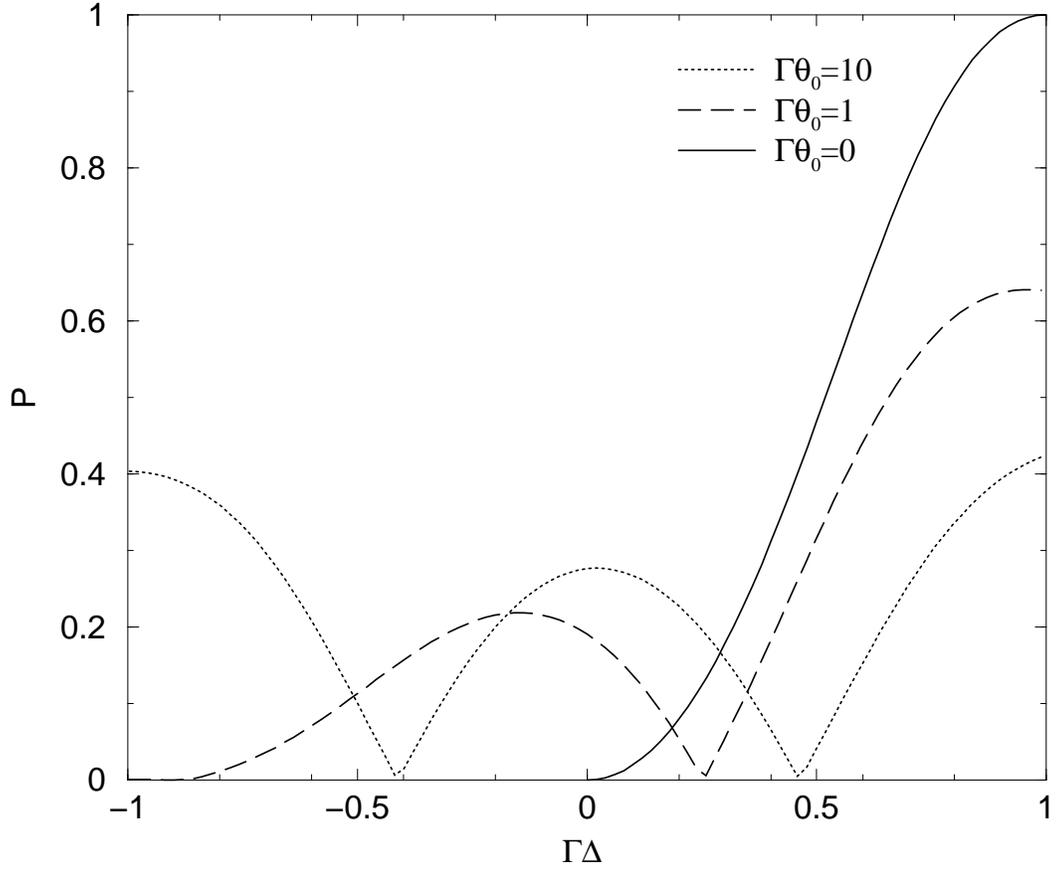}}
\caption{
The polarization degree as a function of
$\Gamma(\sin\theta_0-\sin\theta)$, calculated numerically
using a radiating ring with $a_1=a_2=1$ and highly beamed emission
for different values of $\Gamma\theta_0$, is shown for the purpose of
comparison with the results obtained in the head on approximation
by Eichler and Levinson (2003).}
\label{f4}
\end{figure}
 
\newpage
\begin{figure}[hb]
\centerline{\epsfxsize=140mm\epsfbox{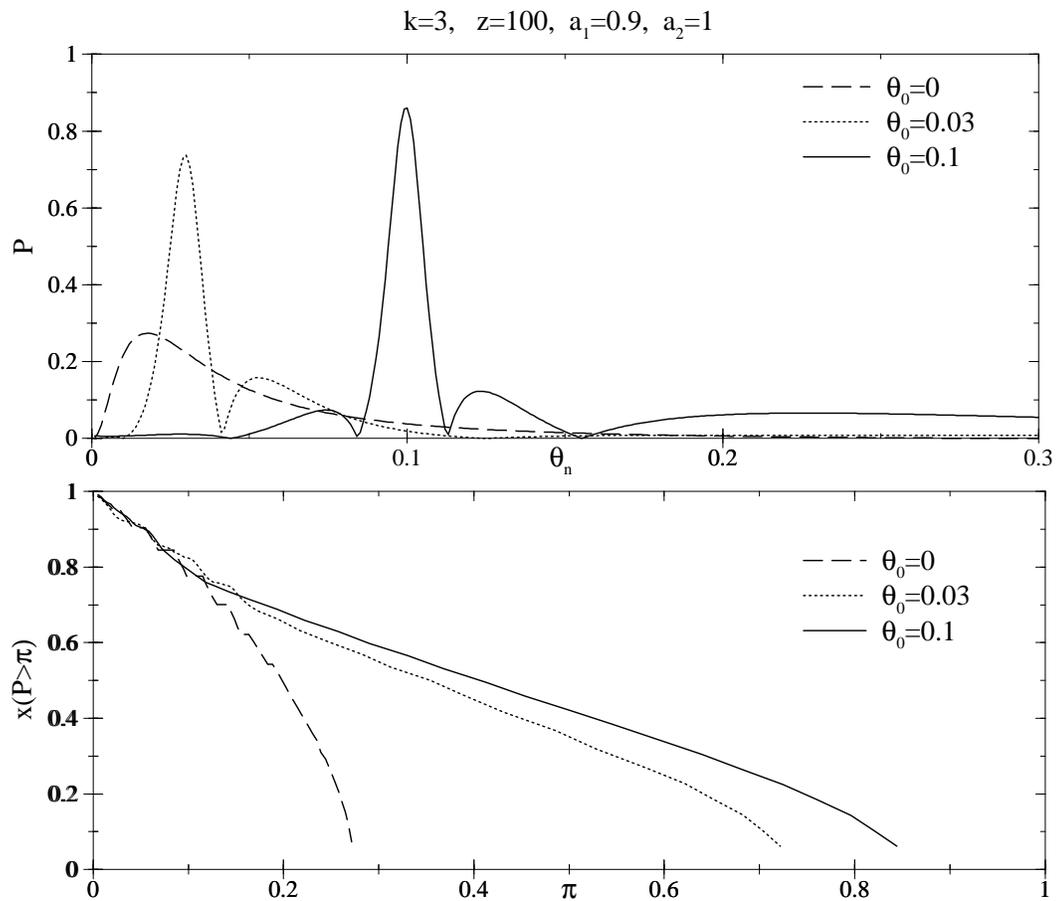}}
\caption{
The upper panel exhibits polarization curves for
different values of the opening angle $\theta_0$.  The
polarization in the inner bump is parallel to the wall, and
changes by 90$^\circ$ across the kinks. The bottom panel displays
the corresponding probability for observing a source with
polarization higher than $\pi$, as given by eq. (\ref{x(P)}) in
the text.}
\label{f5}
\end{figure}
 
\newpage
\begin{figure}[hb]
\centerline{\epsfxsize=140mm\epsfbox{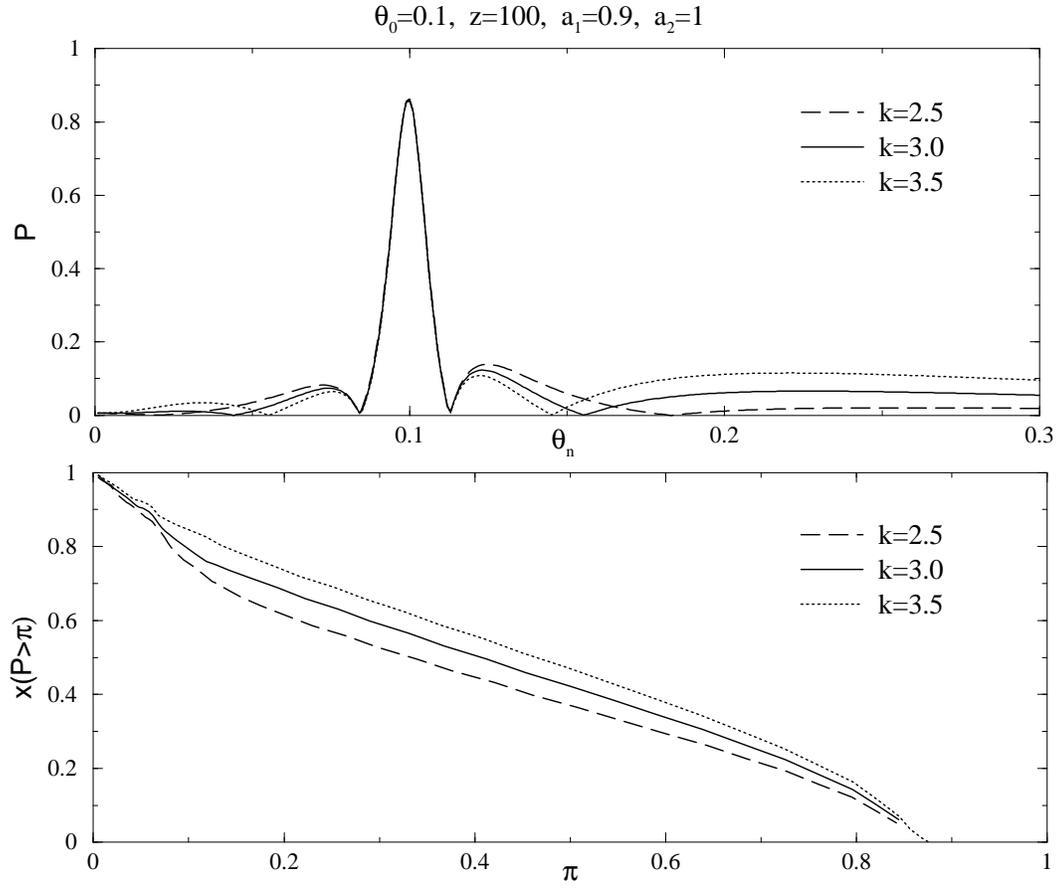}}
\caption{
Same as Fig. 5, but for different values of the index $k$.}
\label{f6}
\end{figure}
 
\newpage
\begin{figure}[hb]
\centerline{\epsfxsize=140mm\epsfbox{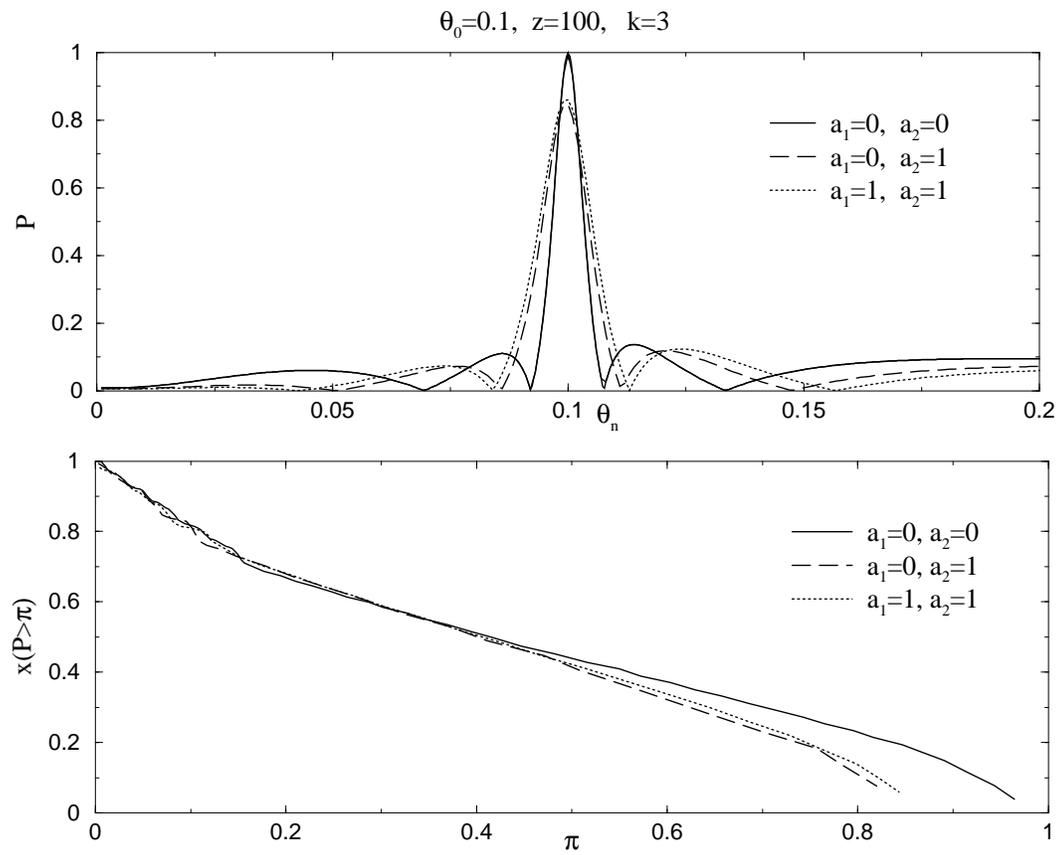}}
\caption{
Same as Fig. 5, but for different ring parameters}
\label{f7}
\end{figure}
 
\newpage
\begin{figure}[hb]
\centerline{\epsfxsize=140mm\epsfbox{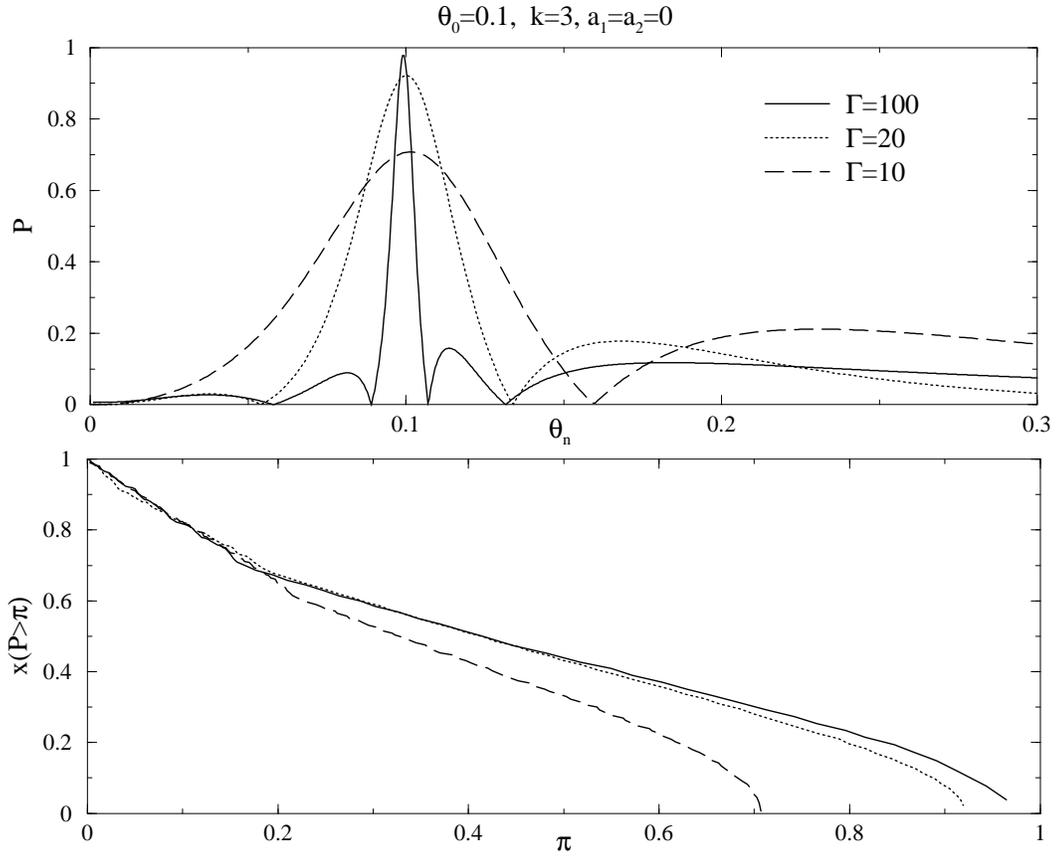}}
\caption{
Dependence of polarization on the Lorentz factor of the Compton sailing
matter. All cases shown were calculated using a point radiation source.}
\label{f8}
\end{figure}
 
\newpage
\begin{figure}[hb]
\centerline{\epsfxsize=140mm\epsfbox{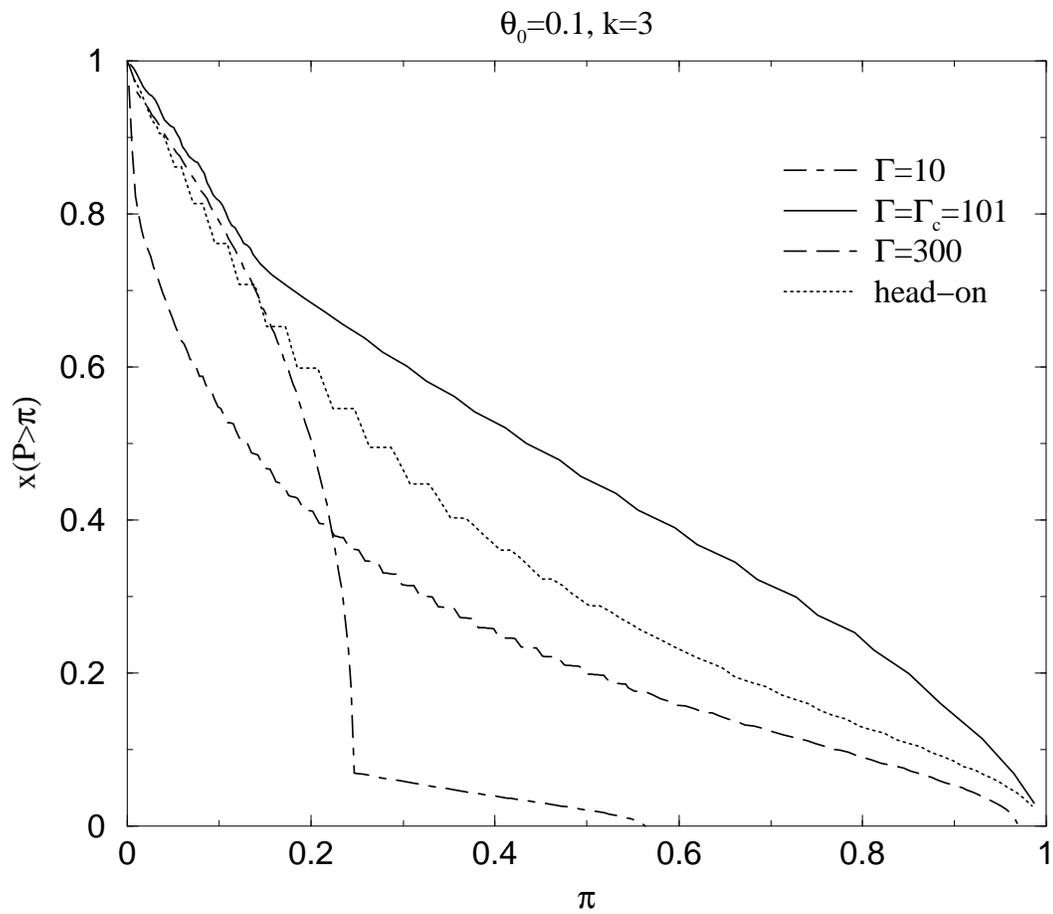}}
\caption{
Comparison of the cumulative probability curves obtained for a
Compton sailing wall (indicated by $\Gamma_c$), with those obtained
in cases where the Lorentz factor of the scattering material deviates from the
Compton sailing value (see text for further discussion).}
\label{f9}
\end{figure}
 
\newpage
\begin{figure}[hb]
\centerline{\epsfxsize=140mm\epsfbox{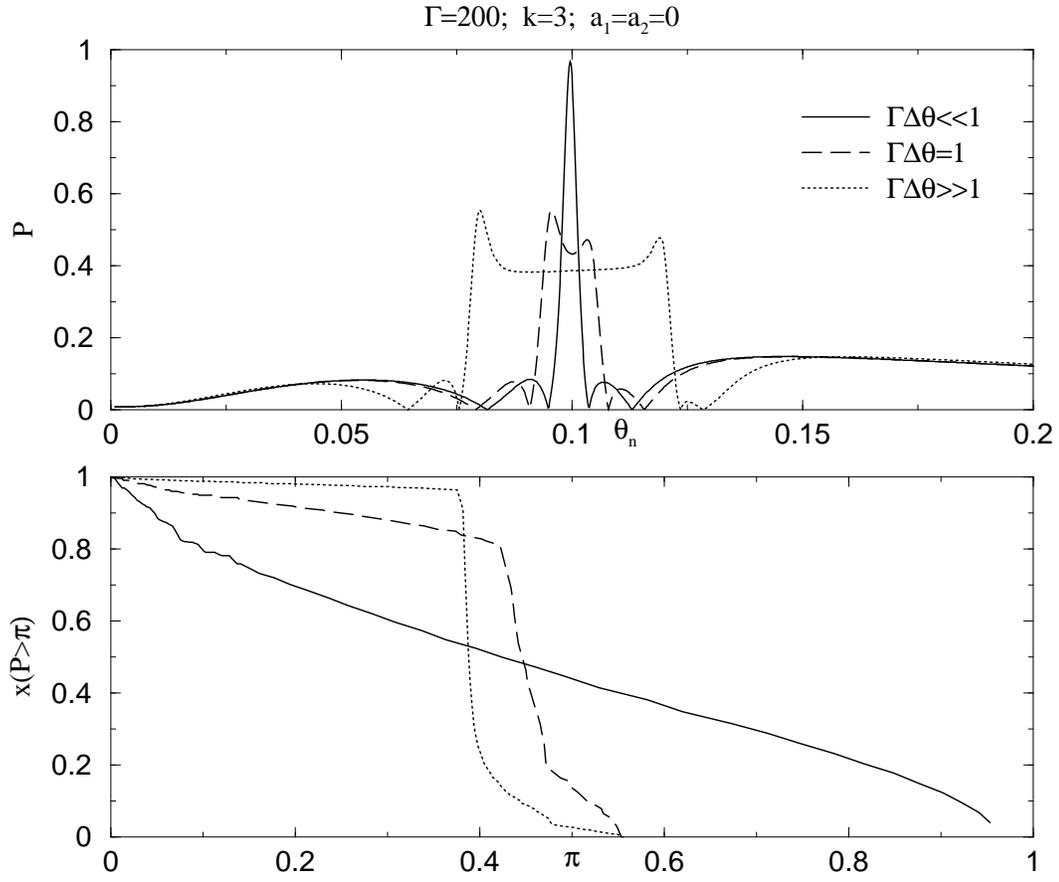}}
\caption{
The polarization produced by scattering off a matter
moving on conical shells encompassing a range of opening angles
between $\theta_0-\Delta\theta$ and $\theta_0+\Delta\theta$.  Each
sub-shell is coasting with a Lorentz factor corresponding to the Compton sailing 
value.  The Loerntz factor of the central sub-shell is indicated..}
\label{f10}
\end{figure}
 
\newpage
\begin{figure}[hb]
\centerline{\epsfxsize=140mm\epsfbox{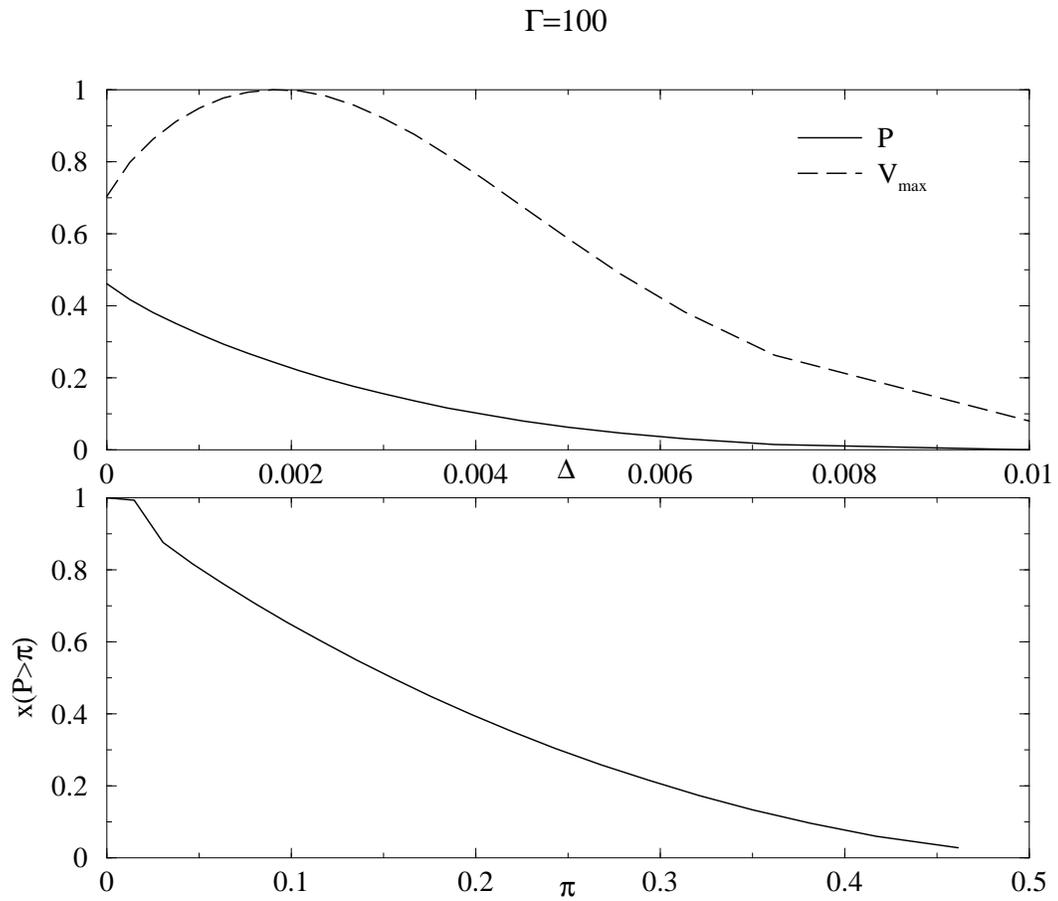}}
\caption{
Effect of multiple scattering inside the wall on the polarization
of emergent radiation.  Shown are the polarization of a
radiation reflected off a semi-infinite, plane-parallel slab, as a function of the
angle $\Delta$ from the wall (upper panel), and the cumulative probability
(lower panel) . The direction of the
incident beam is taken to be normal to wall in the wall's rest frame.}
\label{f11}
\end{figure}

\end{document}